\documentclass[aps,prl,preprint,groupedaddress,showpacs]{revtex4}
\usepackage{graphicx}
\input epsf
\begin{document}

\title{Lightest scalar in the \boldmath{$SU_L(2)\times SU_R(2)$} linear $\sigma$ model
}
\author {
N.N. Achasov}
\author {
G.N. Shestakov}

\affiliation{
   Laboratory of Theoretical Physics,
 Sobolev Institute for Mathematics, Novosibirsk, 630090, Russia}

\date{\today}

\begin{abstract}

We consider the lightest scalar meson in the frame of the
$SU_L(2)\times SU_R(2)$ linear $\sigma$ model, keeping in mind
that this model could be the low energy realization of the
two-flavour QCD. We show that the $\sigma$ field is described by
its four-quark component at least in the $\sigma$ resonance energy
(virtuality) region and the $\sigma\to\gamma\gamma$ decay is the
four quark transition.
  We emphasize that residues of the
$\sigma$ pole in the  $\pi\pi\to\pi\pi$ and
$\gamma\gamma\to\pi\pi$ amplitudes do not give an idea about  the
$\sigma$ meson nature, and the progress in studying the $\sigma$
meson production mechanisms in different processes  could
essentially further us in understanding its nature.

\end{abstract}


\maketitle

The $SU_L(2)\times SU_R(2)$ linear $\sigma$ model \cite{gellman}
\begin{eqnarray}
&&  L=\frac{1}{2}\left
[(\partial_\mu\sigma)^2+(\partial_\mu\overrightarrow{\pi})^2\right
]- \frac{m^2_\sigma}{2}\sigma^2 -
\frac{m^2_\pi}{2}\overrightarrow{\pi}^2 \nonumber\\[6pt] && -\,
\frac{m^2_\sigma - m^2_\pi}{8f^2_\pi}\left [\left (
\sigma^2+\overrightarrow{\pi}\right)^2 + 4f_\pi\,\sigma\left (
\sigma^2+\overrightarrow{\pi}^2\right)\right ]
\end{eqnarray}
had played an outstanding part in the making of hadron physics. In
principle, it could be the low energy realization of the
two-flavour QCD. Up to now this model is an excellent laboratory
for elucidating subtle points in low energy hadron physics. In
particular, we
 showed \cite{annshgn-94}  that in the linear $\sigma$-model there
is a negative background phase which hides the $\sigma$ meson, so
that the  $\pi\pi$  scattering phase shift does not pass over
$90^0$ at putative resonance mass. It has been made clear that
shielding of wide lightest scalar mesons in chiral dynamics is
very natural. This idea was picked up (see, for example, Refs.
\cite{ishida,NNAAVK-2006}) and triggered a new wave of theoretical
and experimental searches for the $\sigma$ and $\kappa$ mesons,
see Particle Data Group Review \cite{pdg-2006}. Below we use our
results \cite{annshgn-94} to analyze the $\sigma$ meson production
in the $\gamma\gamma$ collisions.

 Using the
simplest Dyson equation for the isoscalar scalar $\pi\pi$
scattering amplitude with the real intermediate $\pi\pi$ states
only (in other words, with regard to the real intermediate
$\pi\pi$ states only in every rescattering act) we obtained
\cite{annshgn-94} the simple solution, satisfying both  unitarity
and Adler's self-consistency conditions \cite{false},
\begin{eqnarray}
\label{00amp}
 &&
T^0_0=\frac{T_0^{0(tree)}}{1-i\rho_{\pi\pi}T_0^{0(tree)}}=
\frac{e^{2i\delta^0_0}-1}{2i \rho_{\pi\pi}} = T_{bg}+
e^{2i\delta_{bg}}T_{res}\,,\nonumber\\[6pt] &&
T_0^{0(tree)}=\frac{m_\pi^2-m_\sigma^2}{32\pi f^2_\pi}\left
[5-3\frac{m^2_\sigma-m^2_\pi}{m^2_\sigma-s}-2\frac{m_\sigma^2-m_\pi^2}{s-4m_\pi^2}\ln\left
(1+\frac{s-4m^2_\pi}{m_\sigma^2}\right )\right
]\,,\nonumber\\[6pt] && \delta^0_0=\delta_{bg}+\delta_{res}\,,\ \
\ \rho_{\pi\pi}=\sqrt{1-4m_\pi^2/s}\,,\ \ \
g_{\sigma\pi^+\pi^-}=-\frac{m^2_\sigma- m^2_\pi}{f_\pi}\,,\ \ \
f_\pi=92.4\,\mbox{MeV}\,, \nonumber\\[6pt] &&
T_{res}=\frac{1}{\rho_{\pi\pi}}\left [
\frac{\sqrt{s}\Gamma_{res}(s)}{M^2_{res} - s +
\mbox{Re}\Pi_{res}(M^2_{res})- \Pi_{res}(s)}\right ]
=\frac{e^{2i\delta_{res}}-1}{2i\rho_{\pi\pi}}\,,\ \ \
m_\pi=139.6\,\mbox{MeV}\,, \nonumber\\[6pt] &&
T_{bg}=\frac{\lambda(s)}{1-i\rho_{\pi\pi}\lambda(s)}
=\frac{e^{2i\delta_{bg}}-1}{2i \rho_{\pi\pi}}\,,\ \lambda(s)=
\frac{m_\pi^2-m_\sigma^2}{32\pi f^2_\pi}\left
[5-2\frac{m_\sigma^2-m_\pi^2}{s-4m_\pi^2}\ln\left
(1+\frac{s-4m^2_\pi}{m_\sigma^2}\right )\right
]\,,\nonumber\\[6pt]
 &&
\mbox{Re}\Pi_{res}(s)=-\frac{g_{res}^2(s)}{16\pi}\lambda(s)
\rho_{\pi\pi}^2\,,\ \ \
\mbox{Im}\Pi_{res}(s)=\sqrt{s}\Gamma_{res}(s)=\frac{g_{res}^2(s)}
{16\pi}\rho_{\pi\pi}\,,\nonumber\\[6pt]
  && M^2_{res}= m_\sigma^2 -
\mbox{Re}\Pi_{res}(M^2_{res})\,,\ \ \
g_{res}(s)=\frac{g_{\sigma\pi\pi}}{\left
|1-i\rho_{\pi\pi}\lambda(s)\right |}\,,\ \ \ g_{\sigma\pi\pi}=
\sqrt{\frac{3}{2}}\,g_{\sigma\pi^+\pi^-}\,,
\end{eqnarray}
  where $s$  is the $\pi\pi$ system invariant mass square. These  formulae
show that the resonance contribution is strongly modified by the
chiral background amplitude.

\begin{figure} \centerline{
\epsfxsize=16cm \epsfysize=10cm \epsfbox{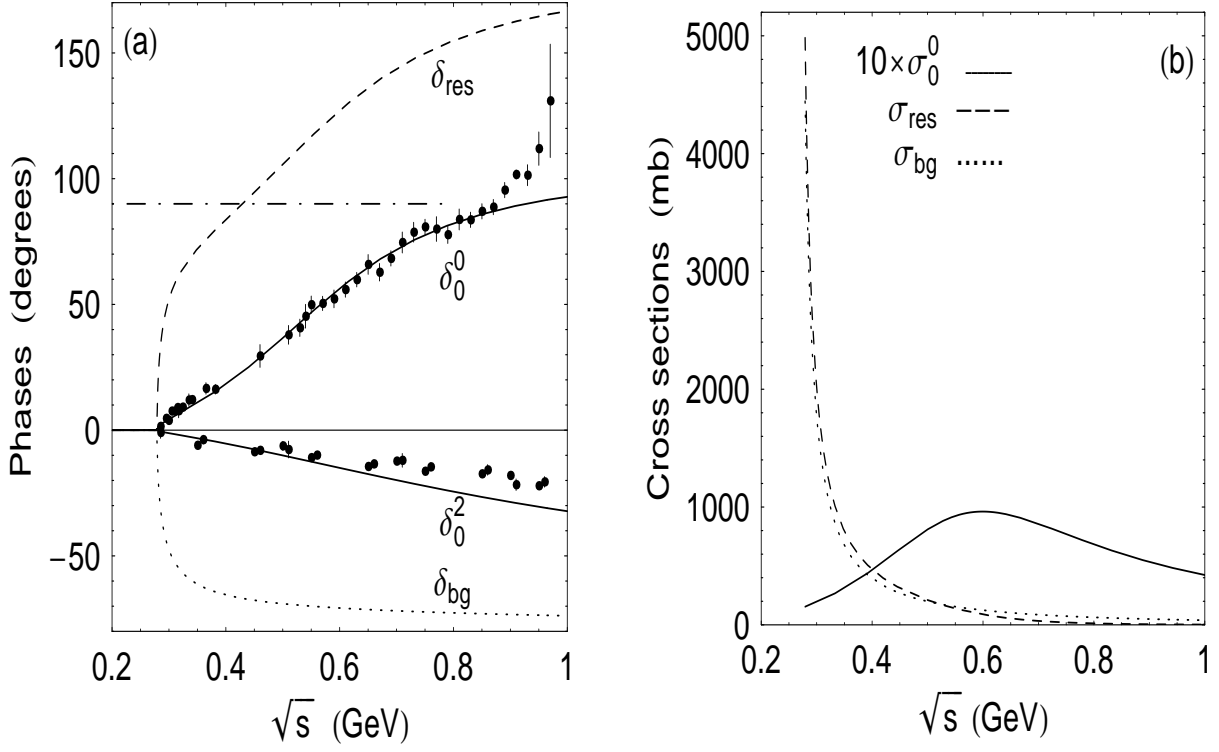}}
 \caption {The $\sigma$ model, our approximation, the details in the text. The data on $\delta^0_0$ and $\delta^2_0$
 from Refs. \cite{HP,DH}, respectively.} \label{fig1}
\end{figure}

In a similar manner we find for the isotensor scalar $\pi\pi$
scattering  amplitude
\begin{eqnarray}
\label{20amp}
 &&
T^2_0=\frac{T_0^{2(tree)}}{1-\imath\rho_{\pi\pi}T_2^{0(tree)}}
=\frac{e^{2i\delta_0^2}-1}{2i\rho_{\pi\pi}},\nonumber\\[6pt] &&
T^{2(tree)}_0 = \frac{m_\pi^2-m_\sigma^2}{32\pi f^2_\pi}\left
[2-2\frac{m_\sigma^2-m_\pi^2}{s-4m_\pi^2}\ln\left
(1+\frac{s-4m^2_\pi}{m_\sigma^2}\right )\right ]\,.
\end{eqnarray}

Our approximation (\ref{00amp}) and (\ref{20amp}) describes the
data acceptably, see Ref. \cite{annshgn-94} and Fig.\ref{fig1}(a),
for the bare $\sigma$ meson mass $m_\sigma = 0.93$ GeV obtained
from fitting $\delta^0_0$ in the region $2m_\pi\leq\sqrt{s} \leq
0.87$ GeV.  The chiral shielding is demonstrated on
Fig.\ref{fig1}(a) with the help of the $\delta_{res}$,
$\delta_{bg}$,  $\delta^0_0$  phases, and on Fig.\ref{fig1}(b)
with the help of the cross sections

\begin{equation}
\label{sections}
 \sigma_{res}=\frac{32\pi}{s}\left |T_{res}\right
|^2 \,,\ \ \ \ \ \ \sigma_{bg}=\frac{32\pi}{s}\left |T_{bg}\right
|^2 \,,\ \ \  \ \ \ \sigma^0_0=\frac{32\pi}{s}\left |T^0_0\right
|^2 \,.
\end{equation}

 Our other results are
\begin{eqnarray}
\label{results}  && M_{res}=0.43\,\mbox{GeV}\,,\ \ \
\Gamma_{res}(M^2_{res})=0.67\,\mbox{GeV}\,,\ \ \ m_\sigma = 0.93\,
\mbox{GeV} \nonumber\\ &&\Gamma^{renorm}_{res}(M^2_{res}) =\left
(1+ d\mbox{Re}\Pi_{res}(s)/ds|_{s=M^2_{res}}\right )^{-1}
\Gamma_{res}(M^2_{res})=0.53\,\mbox{GeV}\,,\nonumber\\[6pt] &&
a^0_0=0.18\, m_\pi^{-1},\   a^2_0=-0.04\, m_\pi^{-1},\
(s_A)^0_0=0.45\, m^2_\pi\,,\   (s_A)^2_0=2.02\, m^2_\pi\,,
\end{eqnarray}
where $a^0_0$ and $a^2_0$ are the scattering lengths, $(s_A)^0_0$
and $(s_A)^2_0$ are the Adler zero positions in $T^0_0$ and
$T^2_0$, respectively.

 The $\sigma$ pole position
\begin{equation}
\label{polpos} s_R=(0.21-i0.26)\times\mbox{GeV}^2,\ \ \
\sqrt{s_R}=M_R-i\Gamma_R/2=(0.52-i0.25)\times\mbox{GeV}\,.
\end{equation}
The residues of the $\sigma$ pole in $T^0_0$ and $T_{res}$,
\begin{eqnarray}
\label{residues} &&
T^0_0\to\frac{g^2_\pi}{s-s_R}\,,\hspace*{3.4cm}
  T_{res}\to \frac{\left ( g_\pi^{res}\right )^2}{s-s_R}\,, \nonumber\\[9pt]
&& g^2_\pi=(0.12+i0.21)\times\mbox{GeV}^2\,,\ \ \ \ \ \ \left (
g_\pi^{res}\right )^2 = - (0.25+i0.11)\times\mbox{GeV}^2
\end{eqnarray}
have large imaginary parts. So, considering the residue of the
$\sigma$ pole in $T^0_0$ as the square of its coupling constant to
the $\pi\pi$ channel is not a clear guide to understanding the
$\sigma$ meson nature.

However the amplitudes on the physical axis (\ref{results}) are
rather significant. Let us consider the propagator of the $\sigma$
field
\begin{equation}
\label{propagator} \frac{1}{D_\sigma (s)}= \frac{1}{M^2_{res} - s
+ \mbox{Re}\Pi_{res}(M^2_{res})- \Pi_{res}(s)}\,,
\end{equation}
which determines $T_{res}$. The contribution to the $\sigma$ meson
self-energy $\Pi_{res}(s)$ is caused by the intermediate $\pi\pi$
states, that is, by the four-quark intermediate state if we keep
in mind that the $SU_L(2)\times SU_R(2)$ linear $\sigma$  model
could be the low energy realization of the two-flavour QCD. This
contribution shifts the  Breit-Wigner (BW) mass greatly $m_\sigma
- M_{res}= 0.50\,\mbox{GeV}$. So, half the BW mass is determined
by the four-quark contribution even if $m_\sigma $ has another
nature, for example, the two-quark one. The imaginary part
dominates  the propagator modulus in the region 300 MeV $<
\sqrt{s}<$ 600 MeV. So, the $\sigma$ field is described by its
four-quark component at least in this energy (virtuality) region
\cite{rho,repino,fari,borz}.

In the field theory approach one has  the following $S$ wave
$\gamma\gamma\to\pi\pi$ amplitudes, satisfying the unitarity
condition, \cite{NNAGNSh-2005}, see also \cite{NNAVVG-1998},
\begin{eqnarray}
\label{ggto+-} && T_S(\gamma\gamma\to\pi^+\pi^-)=
T_S^{Born}(\gamma\gamma\to\pi^+\pi^-) + 8\alpha
I_{\pi^+\pi^-}\,T_S(\pi^+\pi^-\to\pi^+\pi^-)\nonumber\\[3pt] &&
=T_S^{Born}(\gamma\gamma\to\pi^+\pi^-) + 8\alpha
I_{\pi^+\pi^-}\,\left (\frac{2}{3}\,T_0^0 +
\frac{1}{3}\,T_0^2\right )\nonumber\\[3pt]
&&=\frac{2}{3}e^{i\delta^0_0}\left
\{T_S^{Born}(\gamma\gamma\to\pi^+\pi^-)\cos\delta^0_0
  +
8\frac{\alpha}{\rho_{\pi\pi}}\, \mbox{Re}I_{\pi^+\pi^-}
\sin\delta^0_0\right \}\nonumber\\[6pt]
 && + \frac{1}{3}e^{i\delta^2_0}\left
\{T_S^{Born}(\gamma\gamma\to\pi^+\pi^-)\cos\delta^2_0+
8\frac{\alpha}{\rho_{\pi\pi}}\,
\mbox{Re}I_{\pi^+\pi^-}\sin\delta^2_0\right \}\,,
\end{eqnarray}
and
\begin{eqnarray}
\label{ggto00} && T_S(\gamma\gamma\to\pi^0\pi^0)=
 8\alpha
I_{\pi^+\pi^-}\,T_S(\pi^+\pi^-\to\pi^0\pi^0) = 8\alpha
I_{\pi^+\pi^-}\,\left (\frac{2}{3}\,T_0^0 -
\frac{2}{3}\,T_0^2\right )\nonumber\\[3pt]
&&=\frac{2}{3}e^{i\delta^0_0}\left
\{T_S^{Born}(\gamma\gamma\to\pi^+\pi^-)\cos\delta^0_0
  +
8\frac{\alpha}{\rho_{\pi\pi}}\,
\mbox{Re}I_{\pi^+\pi^-}\sin\delta^0_0\right \}\nonumber\\[6pt]
 && - \frac{2}{3}e^{i\delta^2_0}\left
\{T_S^{Born}(\gamma\gamma\to\pi^+\pi^-)\cos\delta^2_0+
8\frac{\alpha}{\rho_{\pi\pi}}\,
\mbox{Re}I_{\pi^+\pi^-}\sin\delta^2_0\right \}\,,
\end{eqnarray}
where
\begin{eqnarray}
\label{I} && I_{\pi^+\pi^-}= \frac{m^2_\pi}{s}\left
(\pi+i\ln\frac{1+\rho_{\pi\pi}} {1-\rho_{\pi\pi}}\right )^2-1\,,\
\ \ s\geq 4m_\pi^2\,,\nonumber\\[9pt]
 &&
T_S^{Born}(\gamma\gamma\to\pi^+\pi^-)=\frac
{8\alpha}{\rho_{\pi^+\pi^-}}\mbox{Im}I_{\pi^+\pi^-}\,.
\end{eqnarray}
Eqs. (\ref{ggto+-}) and (\ref{ggto00}) assume that the $S$ wave
$\pi\pi$ scattering amplitudes lie on the mass shell in the
rescattering loop $\gamma\gamma\to\pi^+\pi^-\to\pi\pi$;
$I_{\pi^+\pi^-}$ is the attribute of the triangle diagram
$\gamma\gamma\to\pi^+\pi^-\to\sigma\, (\mbox{or any scalar})$.

\begin{figure} \centerline{
\epsfxsize=16cm \epsfysize=10cm \epsfbox{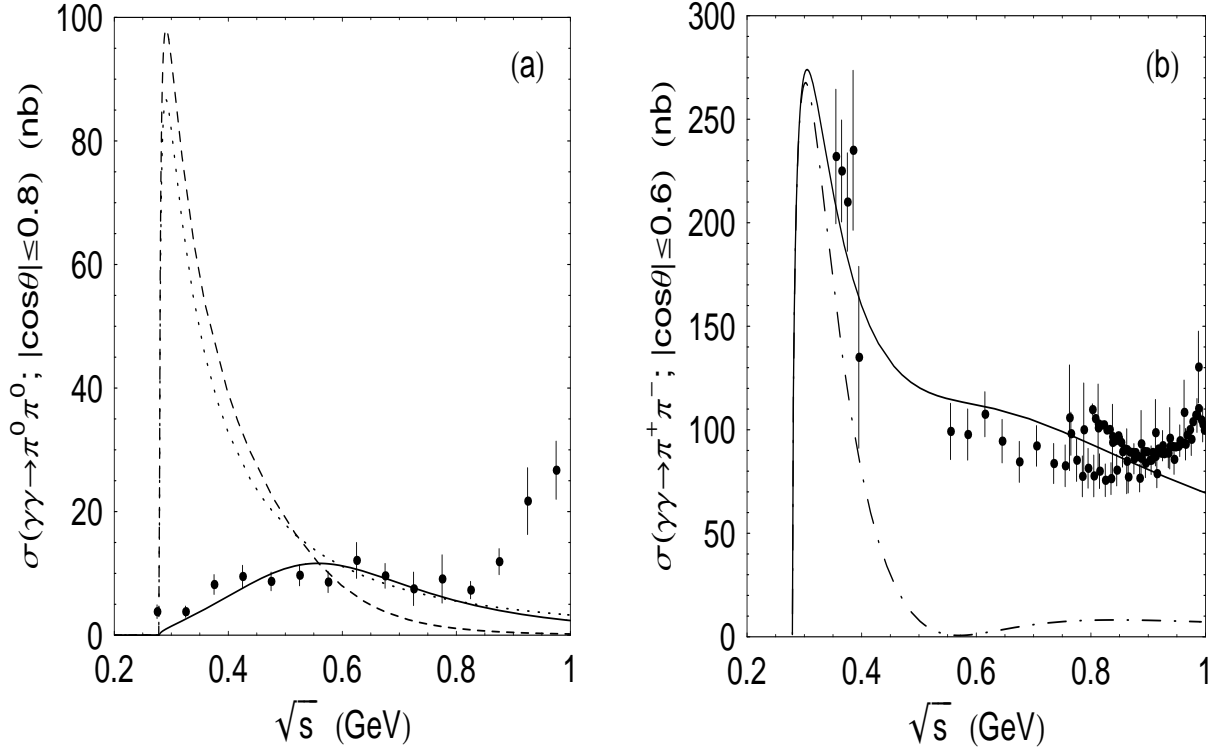}}
 \caption {The $\sigma$ model, our approximation, the details in the text.
 (a) The solid, dashed, and dotted lines correspond to $\sigma_S(\gamma\gamma\to\pi^0\pi^0)$,
$\sigma_{res}(\gamma\gamma\to\pi^0\pi^0)$, and
$\sigma_{bg}(\gamma\gamma\to\pi^0\pi^0)$, respectively. The data
from Ref. \cite{Ma}. (b) The dashed-dotted line corresponds to
$\sigma_S(\gamma\gamma\to\pi^+\pi^-)$. The solid line takes into
account additionally  the  contribution of the higher waves in
$T^{Born}(\gamma\gamma\to\pi^+\pi^-)$ and their interference with
$T_S(\gamma\gamma\to\pi^+\pi^-)$, which takes place for
$|\cos\theta| \leq 0.6$, see details in Ref. \cite{NNAGNSh-2005}.
The data from Refs. \cite{Bo} } \label{fig2}
\end{figure}

The cross sections are
\begin{eqnarray}
\label{cross}
&&
 \sigma_S(\gamma\gamma\to\pi^+\pi^-)=
\frac{\rho_{\pi\pi} }{32\pi s}\left
|T_S(\gamma\gamma\to\pi^+\pi^-)\right |^2\,,\nonumber\\[9pt]
 &&
\sigma_S(\gamma\gamma\to\pi^0\pi^0)= \frac{\rho_{\pi\pi} }{64\pi
s}\left |T_S(\gamma\gamma\to\pi^0\pi^0)\right |^2\,.
\end{eqnarray}

As seen from Fig. \ref{fig2}(a), our calculation agrees
satisfactorily with the $\gamma\gamma\to\pi^0\pi^0$ data in the
$\sigma$ meson region $\sqrt{s} < 0.8$ GeV. Unfortunately, the
$\gamma\gamma\to\pi^+\pi^-$ data are fragmentary in this region,
see Fig. 2(b). Nevertheless, our calculation does not contradict
them.

\begin{figure} \centerline{
\epsfxsize=16cm \epsfysize=10cm \epsfbox{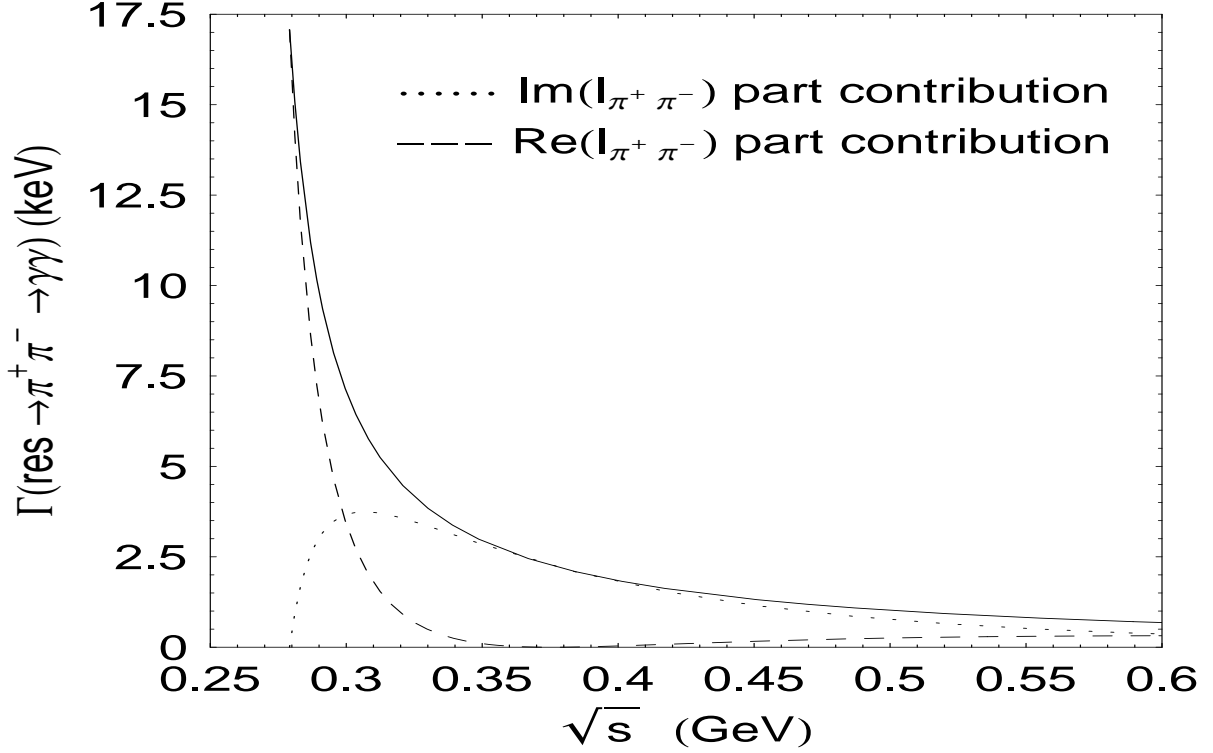}}
 \caption {The $\sigma$ model. The $\sigma\to\gamma\gamma$ decay in our approximation, the details in the text.}
\label{fig3}
\end{figure}

As seen from Eqs. (\ref{ggto+-}) and (\ref{ggto00}) the shielding
of the $\sigma$ meson takes place in the $\gamma\gamma\to\pi\pi$
amplitudes for the strong destructive interference between
$T_{res}$ and $T_{bg}$ as in the $\pi\pi\to\pi\pi$ amplitudes. In
Fig.\ref{fig2}(a) is it demonstrated  with  the help of the cross
sections
\begin{eqnarray}
\label{ggsections}
 &&
 \sigma_{res}(\gamma\gamma\to\pi^0\pi^0)=\frac{\rho_{\pi\pi} }{64\pi
 s}\left |T_{res}(\gamma\gamma\to\pi^0\pi^0)\right |^2 =
\frac{\rho_{\pi\pi} }{64\pi s}\left |8\alpha
I_{\pi^+\pi^-}\times\frac{2}{3}\,T_{res}\right
|^2\,,\nonumber\\[9pt]
 &&
\sigma_{bg}(\gamma\gamma\to\pi^0\pi^0)= \frac{\rho_{\pi^0\pi^0}
}{64\pi
 s}\left |T_{res}(\gamma\gamma\to\pi^0\pi^0)\right |^2=\frac{\rho_{\pi\pi}
}{64\pi s}\left |8\alpha
I_{\pi^+\pi^-}\times\frac{2}{3}\,T_{bg}\right |^2\,,
\end{eqnarray}
and $\sigma_S(\gamma\gamma\to\pi^0\pi^0)$.

Let us consider the $\sigma\to\gamma\gamma$ decay
\begin{eqnarray}
\label{sigmatogg}
&&
 g(res\to\pi^+\pi^-\to\gamma\gamma\,,\,s)=
\frac{\alpha}{2\pi}I_{\pi^+\pi^-}\times
g_{res\,\pi^+\pi^-}(s)\,,\nonumber\\[9pt] &&
\Gamma(res\to\pi^+\pi^-\to\gamma\gamma)=\frac{1}{16\pi\sqrt{s}}\,\left
|g(res\to\pi^+\pi^-\to\gamma\gamma\,,\,s)\right |^2\,,
\end{eqnarray}
where $g_{res\,\pi^+\pi-}(s)=\sqrt{2/3}\times
g_{res}(s)=g_{\sigma\pi^+\pi^-}\Bigl/\Bigl
|1-i\rho_{\pi\pi}\lambda(s)\Bigl |$.

So, the the $\sigma\to\gamma\gamma$ decay is described by the
triangle $\pi^+\pi^-$-loop diagram
$res\to\pi^+\pi^-\to\gamma\gamma$ ($I_{\pi^+\pi^-}$).
Consequently, it is the four-quark transition \cite{a0f0} because
we imply a low energy realization of the two-flavour QCD by means
of the the $SU_L(2)\times SU_R(2)$ linear $\sigma$ model. As Fig.
\ref{fig3} suggests, the real intermediate $\pi^+\pi^-$ state
dominates in $g(res\to\pi^+\pi^-\to\gamma\gamma)$ in the $\sigma$
region $\sqrt{s}< 0.6$ GeV.

 Thus the picture in the physical region is clear and informative.
 But, what  about the pole in the complex $s$-plane? Does the pole
 residue  reveal \cite{mike}  the $\sigma$ indeed?

 Taking the $\gamma\gamma\to\pi^0\pi^0$ amplitude normalization
 like the $T^0_0$ and $T_{res}$ one we obtain
 \begin{eqnarray}
 \label{ggpole}
&&
\frac{1}{16\pi}\sqrt{\frac{3}{2}}\,T_S(\gamma\gamma\to\pi^0\pi^0)\to\frac{g_\gamma
g_\pi}{s-s_R}\,,\hspace*{0.5cm}
\frac{1}{16\pi}\sqrt{\frac{3}{2}}\,T_{res}(\gamma\gamma\to\pi^0\pi^0)\to\frac{g_\gamma^{res}
g_\pi^{res}}{s-s_R}\,,\nonumber\\[9pt] && g_\gamma g_\pi= (-0.45 -
i0.19)\times 10^{-3}\,\mbox{GeV}^2,\hspace*{0.5cm} g_\gamma =
(-0.985+i0.12)\times 10^{-3}\,\mbox{GeV}^2\,, \nonumber\\[9pt] &&
g_\gamma^{res} g_\pi^{res}= (0.53-i0.13)\times
10^{-3}\,\mbox{GeV}^2,\hspace*{0.5cm} g_\gamma^{res}=
(-0.45-i0.95)\times 10^{-3}\,\mbox{GeV}^2,\nonumber\\[9pt]
 &&
 g_\gamma/g_\pi= g_\gamma^{res}/g_\pi^{res}=(-1.61+i1.21)\times
 10^{-3}\,,\nonumber\\[9pt]
 &&
 \Gamma(\sigma\to\gamma\gamma)=|g_\gamma |^2\Bigm/M_R\approx\Gamma_{res}(\sigma\to\gamma\gamma)=
 |g^{res}_\gamma |^2\Bigm/M_R\approx 2\,\mbox{keV}\,.
\end{eqnarray}

It is interesting to compare ratios  $g_\gamma/g_\pi=
g_\gamma^{res}/g_\pi^{res}$ with the ratio
\begin{equation}
 \label{bwratio}
g\left (res\to\pi^+\pi^-\to\gamma\gamma\,,\,M_{res}^2\right
)\Bigm/g_{res}\left(M_{res}^2\right )= (-0.35 + i1.25)\times
10^{-3}\,,
\end{equation}
which are independent on the different normalization in
themselves.

 We find it hard to
believe that anybody could learn the complex but physically clear
dynamics of the $\sigma\to\gamma\gamma$ decay described above from
the residues of Eq. (\ref{ggpole}).

In Ref. \cite{leutwyler} was obtained
\begin{equation}
\label{leutwyler} \sqrt{s_R}= M_R-i\Gamma_R/2 =\left (
441^{+16}_{-8}-i272^{+12.5}_{-9}\right )\times\mbox{MeV}
\end{equation}
with the help of the Roy equation \cite{roy}.

Our result (\ref{polpos}) agrees with the above only
qualitatively. This is natural, because our approximation
(\ref{00amp}) and (\ref{20amp}) gives only a semiquantitative
description of the data at $\sqrt{s}< 0.4$ GeV. In addition, we do
not take into account an effect of the $K\bar K$ channel, the
$f_0(980)$ meson, and so on; that is, do not consider the
$SU_L(3)\times SU_R(3)$ linear $\sigma$ model.

Could the above scenario incorporates the primary lightest scalar
four-quark  state \cite{jaffe}?  Certainly the direct coupling of
this state to $\gamma\gamma$ via neutral vector pairs
($\rho^0\rho^0$ and $\omega\omega$), contained in its wave
function, is negligible $\Gamma(q^2\bar q^2\to
\rho^0\rho^0+\omega\omega\to\gamma\gamma)\approx 10^{-3}$ keV
\cite{1982}. But its coupling to $\pi\pi$ is strong and leads to
$\Gamma(q^2\bar q^2\to\pi^+\pi^-\to\gamma\gamma)$ similar to
$\Gamma(res\to\pi^+\pi^-\to\gamma\gamma)$ in Fig. \ref{fig3}
\cite{f0a0KbarKgg}. Let us add to Eq. (\ref{ggto00})  the
amplitude for the the direct coupling of $\sigma$ to
$\gamma\gamma$  \cite{unitarity}
\begin{equation}
\label{direct}
T_{direct}(\gamma\gamma\to\pi^0\pi^0)=sg^{(0)}_{\sigma\gamma\gamma}g_{res}(s)e^{i\delta_{bg}}\Bigm/D_{res}(s)\,,
\end{equation}
where $g^{(0)}_{\sigma\gamma\gamma}$ is the direct coupling
constant of $\sigma$ to $\gamma\gamma$ , the factor $s$ is caused
by gauge invariance. Fitting the $\gamma\gamma\to\pi^0\pi^0$ data
gives a negligible value  of $g^{(0)}_{\sigma\gamma\gamma}$,
$\Gamma^{(0)}_{\sigma\gamma\gamma}=\left
|M^2_{res}g^{(0)}_{\sigma\gamma\gamma}\right |^2\Bigm/\left (16\pi
M_{res}\right )\approx 0.0034$ keV, in astonishing agreement with
prediction \cite{1982}.

As already noted in Ref. \cite{repino}, the majority of current
investigations of the mass spectra in scalar channels does not
study particle production mechanisms. Because of this, such
investigations are essentially preprocessing experiments, and  the
derivable information is very relative. The progress in
understanding the particle production mechanisms could essentially
further us in revealing the light scalar meson nature. We hope
that it is shown above clearly.

 This work was supported in part by  the Presidential Grant
No. NSh-5362.2006.2 for  Leading Scientific Schools and by the
RFFI Grant No. 07-02-00093 from Russian Foundation for Basic
Research.


\begin{thebibliography}{99}
\bibitem{false}
 Strictly speaking, analitycity is violated in our approximation. The point is that
 $T^0_0$ has a false pole at $s^0_0=0,00137\,m_\pi^2$, but its
residue is extremely small $g^0_0=-0.98\times
10^{-6}\,\mbox{GeV}^2$, $T^0_0\to g^0_0\Bigm/\Bigl ( s -
s^0_0\Bigr )$; please compare with (\ref{residues}).
\bibitem{rho}
  As noted   recently in Ref.  \cite{repino}, even the classic two-quark $\rho$
meson is observed in its resonance region due to the four-quark
component of the $\rho$ meson field.
\bibitem{fari}
 Note also that the authors of Ref. \cite{borz}, using the derivative
$\sigma\pi\pi$ coupling in the $\sigma$ meson propagator, conclude
from the QCD sum-rules that a two-quark current does not describe
the light and broad sigma meson, quite consistent with the present
analysis.
\bibitem{unitarity}
It is evident that the addition of the amplitude (\ref{direct}) to
the amplitude (\ref{ggto00}) does not violate the unitarity
condition.
\end{thebibliography}
\end{document}